\documentclass[aps,prl,showpacs,twocolumn,superscriptaddress]{revtex4}
\usepackage{graphicx} 
\usepackage{xspace}
\usepackage{epsfig}
\usepackage{multirow}
\usepackage{dcolumn}  

\newcommand{\de}{\ensuremath{\Delta E}\xspace}

\newcommand{\br}{\ensuremath{\mathcal{B}}\xspace}

\newcommand{\bb}{\ensuremath{B \overline{B}}\xspace}

\def\myspecial#1{}                   
\def\calL{{\mathcal L}}

\def\Mbc{M_{\rm bc}}

\begin{document}
\myspecial{!userdict begin /bop-hook{gsave 300 50 translate 5 rotate
    /Times-Roman findfont 18 scalefont setfont
    0 0 moveto 0.70 setgray
    (\mySpecialText)
    show grestore}def end}




\title{\quad\\[0.5cm]
 Observation of $B$ Decays to Two Kaons}

\affiliation{Budker Institute of Nuclear Physics, Novosibirsk}
\affiliation{Chiba University, Chiba}
\affiliation{Chonnam National University, Kwangju}
\affiliation{University of Cincinnati, Cincinnati, Ohio 45221}
\affiliation{Justus-Liebig-Universit\"at Gie\ss{}en, Gie\ss{}en}
\affiliation{The Graduate University for Advanced Studies, Hayama, Japan}
\affiliation{University of Hawaii, Honolulu, Hawaii 96822}
\affiliation{High Energy Accelerator Research Organization (KEK), Tsukuba}
\affiliation{Institute of High Energy Physics, Chinese Academy of Sciences, Beijing}
\affiliation{Institute of High Energy Physics, Vienna}
\affiliation{Institute of High Energy Physics, Protvino}
\affiliation{Institute for Theoretical and Experimental Physics, Moscow}
\affiliation{J. Stefan Institute, Ljubljana}
\affiliation{Kanagawa University, Yokohama}
\affiliation{Korea University, Seoul}
\affiliation{Kyungpook National University, Taegu}
\affiliation{Swiss Federal Institute of Technology of Lausanne, EPFL, Lausanne}
\affiliation{University of Ljubljana, Ljubljana}
\affiliation{University of Maribor, Maribor}
\affiliation{University of Melbourne, Victoria}
\affiliation{Nagoya University, Nagoya}
\affiliation{Nara Women's University, Nara}
\affiliation{National Central University, Chung-li}
\affiliation{National United University, Miao Li}
\affiliation{Department of Physics, National Taiwan University, Taipei}
\affiliation{H. Niewodniczanski Institute of Nuclear Physics, Krakow}
\affiliation{Nippon Dental University, Niigata}
\affiliation{Niigata University, Niigata}
\affiliation{University of Nova Gorica, Nova Gorica}
\affiliation{Osaka City University, Osaka}
\affiliation{Osaka University, Osaka}
\affiliation{Panjab University, Chandigarh}
\affiliation{Peking University, Beijing}
\affiliation{RIKEN BNL Research Center, Upton, New York 11973}
\affiliation{University of Science and Technology of China, Hefei}
\affiliation{Seoul National University, Seoul}
\affiliation{Shinshu University, Nagano}
\affiliation{Sungkyunkwan University, Suwon}
\affiliation{University of Sydney, Sydney NSW}
\affiliation{Tata Institute of Fundamental Research, Bombay}
\affiliation{Toho University, Funabashi}
\affiliation{Tohoku Gakuin University, Tagajo}
\affiliation{Tohoku University, Sendai}
\affiliation{Department of Physics, University of Tokyo, Tokyo}
\affiliation{Tokyo Institute of Technology, Tokyo}
\affiliation{Tokyo Metropolitan University, Tokyo}
\affiliation{Tokyo University of Agriculture and Technology, Tokyo}
\affiliation{Virginia Polytechnic Institute and State University, Blacksburg, Virginia 24061}
\affiliation{Yonsei University, Seoul}
 \author{S.-W.~Lin}\affiliation{Department of Physics, National Taiwan University, Taipei} 
 \author{P.~Chang}\affiliation{Department of Physics, National Taiwan University, Taipei} 
 \author{K.~Abe}\affiliation{High Energy Accelerator Research Organization (KEK), Tsukuba} 
 \author{K.~Abe}\affiliation{Tohoku Gakuin University, Tagajo} 
 \author{I.~Adachi}\affiliation{High Energy Accelerator Research Organization (KEK), Tsukuba} 
 \author{H.~Aihara}\affiliation{Department of Physics, University of Tokyo, Tokyo} 
 \author{D.~Anipko}\affiliation{Budker Institute of Nuclear Physics, Novosibirsk} 
 \author{K.~Arinstein}\affiliation{Budker Institute of Nuclear Physics, Novosibirsk} 
 \author{V.~Aulchenko}\affiliation{Budker Institute of Nuclear Physics, Novosibirsk} 
 \author{T.~Aushev}\affiliation{Swiss Federal Institute of Technology of Lausanne, EPFL, Lausanne}\affiliation{Institute for Theoretical and Experimental Physics, Moscow} 
% ...
\collaboration{The Belle Collaboration}

\noaffiliation


\begin{abstract}
 Using 
 449 million $B\overline{B}$ pairs
 collected with the Belle detector at the
KEKB asymmetric-energy $e^+e^-$ collider, we observe clear signals
for $B^+\to \overline{K}{}^0 K^+$ and $B^0 \to \overline{K}{}^0 K^0$ decays with
$5.3\, \sigma$ and $6.0\, \sigma$ significance, respectively. We
measure the branching fractions $\br(B^+\to \overline{K}{}^0 K^+) =
(1.22^{+0.32+0.13}_{-0.28-0.16} )\times 10^{-6}$ and $\br(B^0\to
\overline{K}{}^0 K^0) = (0.87 ^{+0.25}_{-0.20} \pm 0.09)\times
10^{-6}$, and partial-rate asymmetries $A_{CP}(B^+\to
\overline{K}{}^0 K^+) = 0.13^{+0.23}_{-0.24}\pm 0.02$ and
$A_{CP}(B^0\to \overline{K}{}^0 K^0) = -0.58^{+0.73}_{-0.66} \pm 0.04$. 
From a simultaneous fit we also obtain ${\cal B}(B^+ \to K^0 \pi^+
)$ = $(22.8^{+0.8}_{-0.7} \pm 1.3) \times 10^{-6}$ and
$A_{CP}(B^+\to K^0 \pi^+) = 0.03 \pm 0.03 \pm 0.01$. The first and second 
error in 
the branching fractions and the partial-rate asymmetries are statistical and 
systematic, respectively. No signal is
observed for $B^0\to K^+ K^-$ decays, and for this branching fraction we set an 
upper limit of $4.1 \times
10^{-7}$ at the 90\% confidence level.

\end{abstract}


\maketitle

\tighten

{\renewcommand{\thefootnote}{\fnsymbol{footnote}}
\setcounter{footnote}{0}



All $B\to K\pi,\ \pi\pi$ decays have now been
observed~\cite{br,prl89,prl91,prl91b}, and direct $CP$ violation
has been established in $B^0\to K^+\pi^-$~\cite{acp,prl93}. The
measurements of these hadronic $b \to s$ and $b \to u$ transitions
have provided essential information for our understanding of $B$
decay mechanisms, and are probes for possible new
physics.
What remains are the $B\to \overline{K} K$ modes, for which measurements with
good accuracy are needed.
Some measurements for these modes were reported by the Belle and BaBar 
collaborations \cite{aub1,kk}.

In this Letter we report the  observation of $B^0 \to
\overline{K}{}^0 K^0$ and $B^+ \to \overline{K}{}^0 K^+$ decays \cite{naub1}. These
decays are expected to be dominated by the loop-induced $b\to
d\overline ss$ process (called a $b\to d$ penguins) shown in
Fig.~\ref{fig:kkfeynman}(a). When compared with the $b\to s\overline{d}d$ 
penguin-dominated $B^0\to K^0\pi^+$ decay,
these modes are expected to be suppressed by a factor of roughly
1/20, with branching fractions expected at the $10^{-6}$
level~\cite{plb253,plb341}. 
We also
search for $B^0\to K^+ K^-$, which, at lowest order, arises from a $b \to u$
$W$-exchange process (Fig. \ref{fig:kkfeynman}(b)) or from final-state
interactions~\cite{aj}.

\begin{figure}[b!]
\includegraphics[width=0.25\textwidth]{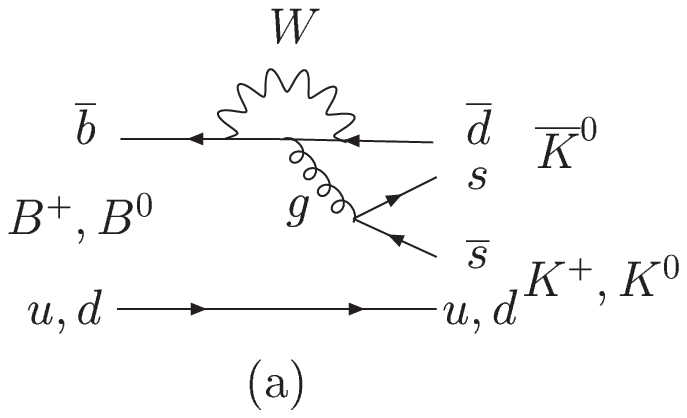}
\includegraphics[width=0.22\textwidth]{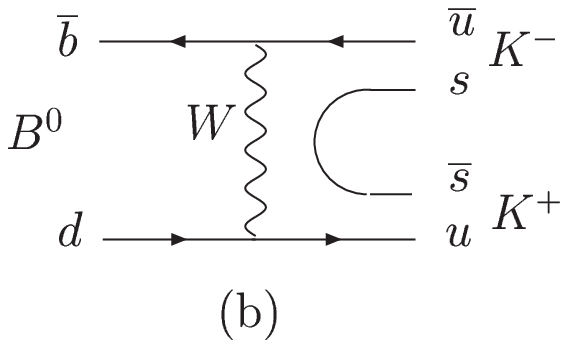}
\caption{The $b \to d$ penguin diagram (a) for $B^+ \to
\overline{K}{}^0 K^+$ and $B^0 \to \overline{K}{}^0 K^0$ modes, and $b
\to u$ $W$-exchange diagram (b) for $B^0 \to K^+ K^-$ decay.}
\label{fig:kkfeynman}
\end{figure}

For the decay modes with significant signal, we also measure the
partial-rate asymmetry,
\begin{equation}
 A_{CP} \equiv \frac{N(\overline{B}\to \overline{f}) - N(B\to f)}
          {N(\overline{B}\to \overline{f}) +N(B\to f)},
\end{equation}
where $f$ denotes $\overline{K}{}^0 K^0$, $\overline{K}{}^0 K^+$ or
$K^0 \pi^+$. Direct $CP$ violation is expected to be sizable in
$B^0 \to \overline{K}{}^0 K^0$ and $B^+ \to \overline{K}{}^0 K^+$
decays~\cite{plb253}, while mixing-dependent $CP$ violation can be
measured in $B^0 \to \overline{K}{}^0 K^0 $ (and
$K^+K^-$)~\cite{plb341}.

%

This analysis is based on a sample of (449.3 $\pm$ 5.7) $\times 10^6$ $\bb$
pairs collected with the Belle detector at the KEKB $e^+e^-$ asymmetric-energy
(3.5 on 8~GeV) collider~\cite{kur}. 
The production rates of $B^+B^-$ and $B^0\overline{B}{}^0$ pairs are assumed to
be equal. Throughout this paper, the inclusion of the charge-conjugate decay is implied, unless explicitly stated otherwise.

The Belle detector is a large-solid-angle magnetic
spectrometer that consists of a silicon vertex detector (SVD),
a 50-layer central drift chamber (CDC), an array of
aerogel threshold Cherenkov counters (ACC),
a barrel-like arrangement of time-of-flight
scintillation counters (TOF), and an electromagnetic calorimeter
comprised of CsI(Tl) crystals located inside
a superconducting solenoid coil that provides a 1.5~T
magnetic field.  An iron flux-return located outside
the coil is instrumented to detect $K_L^0$ mesons and to identify
muons.  The detector is described in detail elsewhere~\cite{aba}.
Two different inner detector configurations were used. For the first sample
of 152 million $\bb$ pairs (Set I), a 2.0 cm radius beampipe
and a 3-layer silicon vertex detector were used;
for the latter  297 million $\bb$ pairs (Set II),
a 1.5 cm radius beampipe, a 4-layer silicon detector
and a small-cell inner drift chamber were used \cite{svd2}.

Primary charged tracks are required to have a distance of closest approach
to the interaction point (IP) of less than 4 cm in the beam
direction ($z$) and less than 0.1 cm in  the transverse plane.
Charged kaons and pions are identified using $dE/dx$ information
from the CDC and Cherenkov light yields in the ACC, which are
combined to form a $K$-$\pi$ likelihood ratio, $\mathcal{R}(K/\pi)
= \mathcal{L}_K/(\mathcal{L}_K+\mathcal{L}_\pi)$, where
$\mathcal{L}_{K}$ $(\mathcal{L}_{\pi})$ is the likelihood that the
track is a kaon (pion). Charged tracks with
$\mathcal{R}(K/\pi)>0.6$ ($<0.4$) are regarded as kaons (pions) for $B^+\to
\overline{K}{}^0 K^+$ ($B^+ \to K^0 \pi^+$) decays. A tighter requirement,
${\mathcal R}(K/\pi) > 0.9$, is used for the $B^0 \to K^+ K^-$
selection due to the large background from $B^0 \to K^+ \pi^-$.
 Furthermore, in all decay modes we reject charged tracks consistent with an
electron hypothesis.

Candidate $K^0$ mesons are observed as $K^0_S \to \pi^+ \pi^-$
decay with the branching fraction taken from Ref.~\cite{pdg}. We pair oppositely-charged tracks assuming the pion hypothesis
and require the invariant mass of the
pair to be within $\pm 18$ MeV/$c^2$ of the nominal $K_S^0$ mass. The
intersection point of the $\pi^+ \pi^-$ pair must be displaced from the IP.

Two variables are used to identify $B$ candidates: the beam-energy constrained mass,
$M_{\rm bc} \equiv
\sqrt{E^{*2}_{\mbox{\scriptsize beam}}/c^4 - p_B^{*2}/c^2}$, and the energy difference,
$\Delta E \equiv E_B^* - E^*_{\mbox{\scriptsize beam}}$, where
$E^*_{\mbox{\scriptsize beam}}$ is the run-dependent beam energy and
$E^*_B$ and $p^*_B$ are
the reconstructed energy and momentum of the $B$ candidates in the
center-of-mass (CM) frame, respectively. Events with
$M_{\rm bc} > 5.20$ GeV/$c^2$ and $|\Delta E| < 0.3~{\rm GeV}$
are selected for the analysis.

The dominant background is from $e^+e^- \to q\overline q ~(
q=u,d,s,c )$ continuum events. We use event topology to
distinguish the spherically distributed $B\overline{B}$
events from jet-like continuum background. We combine a set of
modified Fox-Wolfram moments \cite{pi0pi0} into a Fisher
discriminant. Signal and background likelihoods are formed, based on a
GEANT-based~\cite{geant} Monte Carlo (MC) simulation, from the
product of the probability density function (PDF) for the Fisher
discriminant and that for the cosine of the angle between the $B$
flight direction and the positron beam. Suppression of continuum
is achieved by applying a requirement on a likelihood ratio
$\mathcal{R} = {\calL}_{\rm sig}/({\calL}_{\rm sig} + {\calL}_{q
\overline{q}})$, where ${\cal L}_{\rm sig}$ (${\mathcal L}_{q
\overline{q}}$) is the signal ($q \overline{q}$) likelihood.
Additional continuum background suppression is achieved through use of a
$B$-flavor tagging algorithm \cite{tagging}, which provides a discrete
variable indicating the flavor of the tagging $B$ meson and a quality
parameter $r$, with continuous values ranging from 0 (for no
flavor-tagging information) to 1 (for unambiguous flavor assignment).
Events with a high value of $r$ are considered well-tagged and hence are
unlikely to have originated from continuum processes.
We classify events as well-tagged ($ r>  0.5$) and poorly-tagged ($r\leq 0.5$), and for each category of Set I and Set II we determine a continuum suppression
requirement for ${\cal R}$ that maximizes the value of
$N_{\rm sig}^{\rm exp}/\sqrt{N_{\rm sig}^{\rm exp}+N_{q
\overline{q}}^{\rm exp}}$. Here, $N_{\rm sig}^{\rm exp}$ denotes the
expected signal yields based on MC simulation and the branching fractions of the
previous measurement \cite{kk}, and
$N_{q\overline{q}}^{\rm exp}$ denotes the expected $q\overline q$
yields as estimated from sideband data ($M_{\rm bc}<5.26$ GeV/$c^2$ and $|\Delta E|<$ 0.3 GeV).

Background contributions from other $\Upsilon(4S) \to B\overline B$
events are investigated with a large MC sample that includes
events from $b\to c$ transitions and charmless $B$ decays. After
all the selection requirements, no $B\overline{B}$ background is
found for the $B^0 \to \overline{K}{}^0 K^0$ mode. Due to $K$-$\pi$
misidentification, large $B^0\to K^+\pi^-$ and $B^+\to K^0 \pi^+$
feed-across backgrounds appear in the $B^0\to K^+ K^-$ and $B^+\to
\overline{K}{}^0 K^+$ samples, respectively.
At low $\Delta E$ values, a small background contribution from 
charmless $B$ decays --- mainly from $B^+ \to K^+ \pi^0 \overline{K}{}^0$
($\overline{B}{}^0\to K^{*-} \pi^+$) --- is found for the
$B^+ \to \overline{K}{}^0 K^+$ ($B^+ \to K^0 \pi^+$) decay mode. However,
experimentally the 
$B^+ \to K^+ \pi^0 \overline{K}{}^0$ mode only has an upper limit.

The signal yields are extracted by performing extended unbinned two-dimensional
maximum likelihood (ML) fits to the ($M_{\rm bc}$, $\Delta E$)
distributions.
The likelihood for each mode is defined as
\begin{eqnarray}
\mathcal{L} & = & \exp (-\sum_{l,k,j} N_{l,k,j}) \prod_i
(\sum_{l,k,j} N_{l,k,j} {\mathcal P}^i_{l,k,j}), \;\;\;
\\
\mathcal{P}^i_{l,k,j} & = & \frac{1}{2}[1-q^i\cdot A_{CPj}]\,
P_{l,k,j}(M^i_{{\rm bc}}, \Delta E^i),
\end{eqnarray}
where $i$ is the event identifier, $l$ indicates Set I or Set II,
$k$ distinguishes the two $r$ categories and $j$ runs
over all components included in the fitting function --- one for
the signal and the others for continuum, feed-across and charmless
$B$ backgrounds. 
$N_{l,k,j}$ represent the number of
events, and $P_{l,k,j}(M^i_{{\rm bc}}$, $\Delta E^i)$ are the two-dimensional 
PDFs, which are the
same in the two $r$ categories for all fit components except for the
continuum background. The parameter $q$ indicates the $B$-meson flavor: $q=+1\
(-1)$ for $B^+$ and $B^0$ ($B^-$ and $\overline{B}{}^0$). Unlike
$\overline{K}{}^0 K^+$, the $\overline{K}{}^0 K^0$ and $K^+K^-$
channels are not self-tagged and the $B$ meson flavor must be
determined from the other $B$. To account for the effect of
$B\overline{B}$ mixing and imperfect tagging, the term $A_{CP}$
for the signal in Eq.~(3) has to be replaced by
$A_{CP}(1-2\chi_d)(1-2w_k)$, where $\chi_d$ = 0.188 $\pm$ 0.003
\cite{pdg} is the time-integrated mixing parameter. The $\chi_d$ value of continuum events is set to zero. The wrong-tag
fraction $w_k$, which depends on the value of $r$, is
determined from a high statistics sample of self-tagged $B^0 \to
D^{*-}\pi^+, D^{*-}\rho^+$ and $D^{(*)-}l^+\nu_l$
events~\cite{tagging}.

\begin{table*}[t!]
\begin{center}
\caption{Fitted signal yields, product of
 efficiencies and sub-decay branching fractions $({\cal B}_s)$, branching
 fractions, significance ($\Sigma$), and partial-rate asymmetries for individual modes. The first and second error in 
the branching fractions and the partial-rate asymmetries are statistical and 
systematic, respectively.}
\begin{tabular}{lccccc}
\hline\hline
~Mode~ & Yield & Eff.$\times {\cal B}_s$(\%) & ${\cal B}(10^{-6})$& $\Sigma(\sigma)$ & $A_{CP}$\\
\hline ~$K^+K^-$ & $ 2.5^{+5.0}_{-3.7}$ & 6.18 &
 $0.09^{+0.18}_{-0.13}\pm 0.01\ (<0.41)$& 0.6 & -\\
~$\overline{K}{}^0K^+$ & $36.6^{+9.7}_{-8.3}$ & $6.72$
 & $1.22^{+0.32+0.13}_{-0.28-0.16}$ & 5.3 & $0.13^{+0.23}_{-0.24}\pm 0.02$\\
~$K^0 \pi^+$ & $1252^{+41}_{-39}$ & 12.21 &
 $22.8^{+0.8}_{-0.7} \pm 1.3$ & 53.1 & $0.03 \pm 0.03 \pm 0.01$ \\
~$\overline{K}{}^0 K^0$ & $23.0^{+6.5}_{-5.4}$ & $5.89$ &
 $0.87^{+0.25}_{-0.20} \pm 0.09 $ & 6.0 & $-0.58^{+0.73}_{-0.66}\pm 0.04$ \\
\hline\hline
\end{tabular}
\label{tab:kk}
\end{center}
\end{table*}

All the signal PDFs ($P_{l,k,j=\mathrm{signal}}(M_{\rm bc},\Delta
E)$) are parameterized by smoothed two-dimensional histograms
obtained from correctly reconstructed signal MC based on the Set I
and Set II detector configurations. Signal MC events are generated
with the PHOTOS~\cite{photos} simulation package to take into
account final-state radiation. Since the $M_{\rm bc}$ signal
distribution is dominated by the beam-energy spread, we apply small 
corrections to the signal peak position and resolution determined using 
$B^+ \to \overline{D}{}^0\pi^+$ from data ($\overline{D}{}^0 \to K^0_S\pi^+
\pi^-$ is used for the $\overline{K}{}^0 K^0$ mode, while
$\overline{D}{}^0\to K^+\pi^-$ is used for the other three modes)
with small mode-dependent corrections obtained from MC. 
The resolution for the $\Delta E$ distribution is calibrated
using the invariant mass distributions of high momentum ($p_{\rm Lab}> 3$
GeV/$c$)  $D$ mesons. The decay mode
$\overline{D}{}^0\to K^+\pi^-$ is used for $B^0\to K^+ K^-$,
$D^+\to K^0_S \pi^+$ for $B^+\to K^0\pi^+$ and
$\overline{D}{}^0\to K^0_S\pi^+\pi^-$ for $B^0\to \overline{K}{}^0
K^0$.

The continuum background PDF is described by a product of a linear
function for $\Delta E$ and an ARGUS function, $f(x) = x
\sqrt{1-x^2}\,\exp[ -\xi (1-x^2)]$, where $x$ = 
$\Mbc c^2$/$E^*_{\rm beam}$ \cite{argus}. The overall normalization, $\Delta
E$ slope and ARGUS parameter $\xi$ are free parameters in the fit.
The background PDFs for charmless $B$ decays for the
$K^0 \pi^+$ and $\overline{K}{}^0K^+$ modes are both modeled by
smoothed two-dimensional histograms obtained from a large MC sample.
We also use smoothed two-dimensional histograms to describe the 
feed-across backgrounds for the $K^+ K^-$ ($\overline{K}{}^0K^+$) mode, since
the background $K^+\pi^-$ ($K^0\pi^+$) events have ($\Mbc, \de$) shapes similar
to the signal, except for the $\de$ peak positions shifted by $\simeq45$ MeV.
We perform a simultaneous fit
for $B^+ \to \overline{K}{}^0 K^+$ and $B^+ \to K^0 \pi^+$, since
these two decay modes feed into each other. Because the branching 
fraction of $B^0 \to K^+K^-$ is small, we also treat the yields of $K^+\pi^-$ 
feed-across events as free parameters in the fit.

When likelihood fits are performed, all the $N_{l,k,j}$ are allowed to float except
for the feed-across backgrounds in the $\overline{K}{}^0K^+$ and $K^0\pi^+$ modes.
The $M_{\rm bc}$ and $\Delta E$ projections of the fits are
shown in Fig.~\ref{fig:kk}.
The branching fraction in each mode is calculated by dividing the 
efficiency-corrected total signal yield by the number of $B\overline{B}$
pairs. In Table \ref{tab:kk}, a sum of fitted signal yields and the average
efficiency are listed.

The fitting systematic errors include the signal PDF modeling,
which we estimate from the deviations after varying each parameter of the
signal PDFs by one standard deviation in the calibration factors, and the modeling of the
charmless $B$ background. Since the $\Delta E$ values of the charmless $B$ 
events are typically less than $-0.12$ GeV, the systematic error due to the 
modeling of the charmless $B$ background is evaluated by requiring
$\Delta E>-0.12$ GeV.
 At each step, the yield deviation is
added in quadrature to provide the fitting systematic errors, and the
statistical significance is computed by taking the square root of the
difference between the value of  $-2\ln\mathcal L$  for the best fit and
that for zero signal yield. 
For $B^+ \to \overline{K}{}^0 K^+$, systematic uncertainty is included in the
significance calculation by varying the feed-across background (which is the 
dominant uncertainty) by $1 \sigma$ in the direction that lowers the 
significance. For the other decay modes, the effect of systematic uncertainty
on the significance is negligible.

\begin{table}[t!]
\begin{center}
\caption{Summary of systematic errors, given in percent. }
\begin{tabular}{lccccc}
\hline\hline & $K^+ K^-$ & $\overline{K}{}^0 K^+$ & $K^0 \pi^+$ &
$\overline{K}{}^0 K^0$ \cr 
\hline 
Signal PDF & $^{+1.3}_{-1.4}$ & $\pm$0.2 & $\pm$0.2 & $^{+0.5}_{-0.6}$ \cr 
Charmless $B$ background & 0.0 & $-0.9$ & $-0.1$ & 0.0 \cr
$\mathcal{R}$ requirement & $\pm$0.8 & $\pm$1.4 & $\pm$1.1 & $\pm$3.3 \cr
Tracking & $\pm$2.0 & $\pm$1.0 & $\pm$1.0 & 0.0 \cr 
$\mathcal{R}(K/\pi)$ requirement & $\pm$3.9 & $\pm$1.5 & $\pm$1.3 & 0.0 \cr 
$K^0_S$ reconstruction & 0.0 & $\pm$4.9 & $\pm$4.9 & $\pm$9.8 \cr 
\# of feed-across & 0.0 & $^{+9.4}_{-11.9}$ & $^{+0.2}_{-0.4}$ & 0.0 \cr 
\# of $B\overline{B}$ & $\pm$1.3 & $\pm$1.3 & $\pm$1.3 & $\pm$1.3 \cr
Signal MC statistics& $\pm$1.1 & $\pm$1.1 & $\pm$0.8 & $\pm$0.8 \cr
\hline 
Sum & $^{+4.9}_{-5.0}$ & $^{+11.0}_{-13.2}$ & $\pm$5.5 & $\pm$10.5 \cr 
\hline \hline
\end{tabular}
\label{tab:sys}
\end{center}
\end{table}
\begin{figure}[h!]
\includegraphics[width=0.48\textwidth]{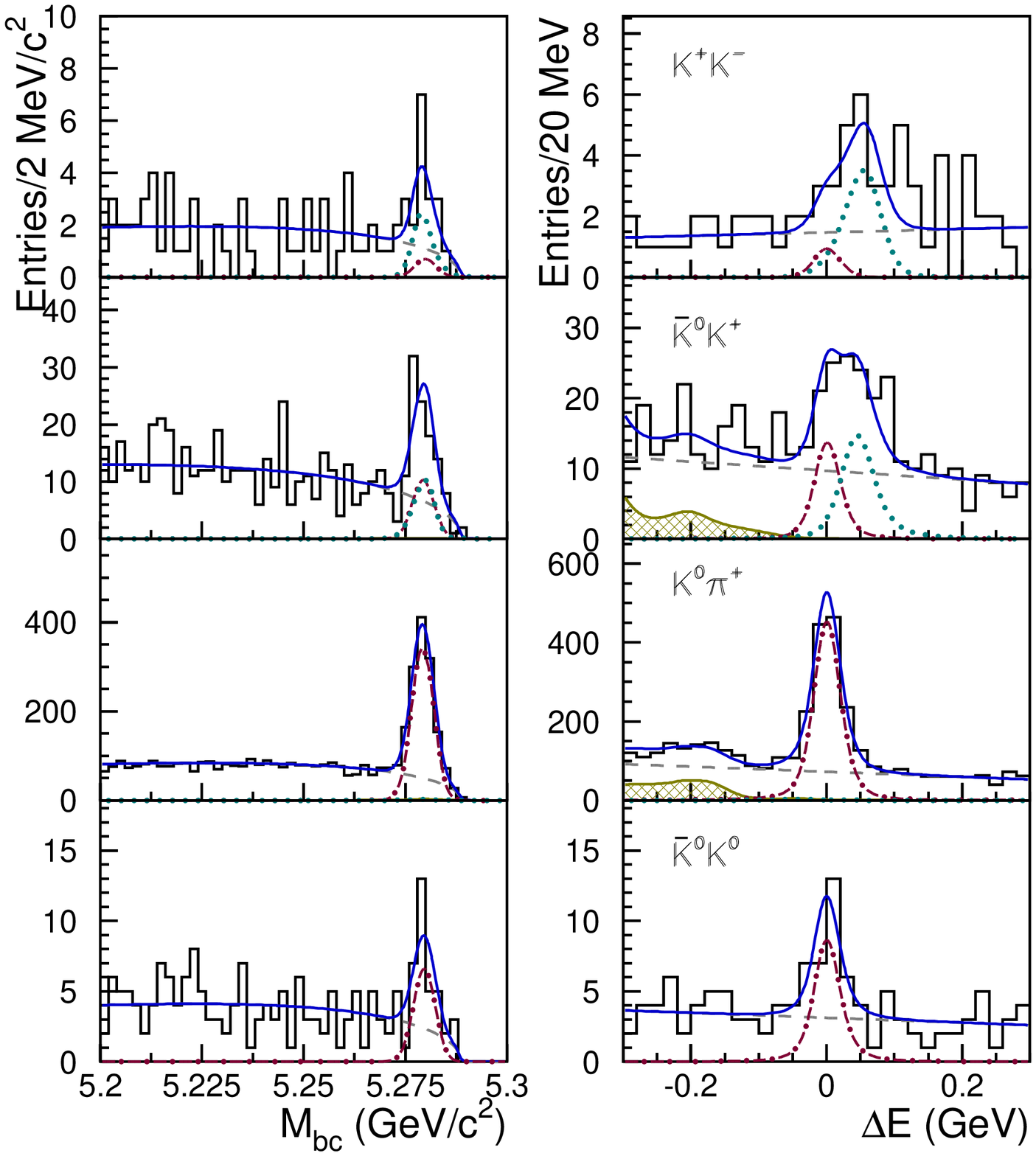}
\caption{$M_{\rm bc}$ (left) and $\Delta E$ (right) distributions
for $B\to K^+K^-$, $\overline{K}{}^0 K^+$, $K^0 \pi^+$ and
$\overline{K}{}^0 K^0$ candidates. The histograms show the data,
while the curves represent the various components from the fit:
signal (dot-dashed), continuum (dashed), charmless $B$
decays (hatched), feed-across background from misidentification
(dotted), and sum of all components (solid). The $\Mbc$ and
$\de$ projections of the fits are for events that have
$|\de|<0.06$ GeV (left) and $5.271$ GeV/$c^2 < {M_{\rm bc}} <
5.289$ GeV/$c^2$ (right), respectively.} \label{fig:kk}
\end{figure}

The MC-data efficiency difference due to the requirement on the
likelihood ratio $\mathcal{R}$ is investigated using the $B^+\to
\overline{D}{}^0\pi^+ $ ($\overline{D}{}^0\to K^0_S\pi^+\pi^-$ for
$\overline{K}{}^0K^0$ and $\overline{D}{}^0\to K^+\pi^-$ for the others)
samples. The systematic
error due to the charged-track reconstruction efficiency is estimated
to be $1$\% per track using partially reconstructed $D^*$
events. The systematic error due to ${\cal R}(K/\pi)$ selection is
1.3\% for pions and 1.5\% for kaons, respectively. Due to
the tighter ${\cal R}(K/\pi)$ selection of kaons in the $K^+K^-$ mode,
the assigned systematic uncertainty is 1.9\% per kaon track. The
$K_S^0$ reconstruction is verified by comparing the
ratio of $D^+\to K_S^0\pi^+$ and $D^+\to K^-\pi^+\pi^+$ yields
with the MC expectation. 
 We vary the yields of feed-across background by $\pm 1 \sigma$ to check the
effect from the constraint on the feed-across background. Possible systematic 
uncertainties due to the description of final-state radiation have been studied 
by comparing the latest theoretical calculations with the PHOTOS 
MC \cite{photoserr}. These uncertainties were found to be negligible and thus 
no systematic error is assigned due to PHOTOS. 
The systematic error due to the uncertainty in the total number of $\bb$ pairs 
is 1.3\% and the error due to signal MC statistics is in the range 0.8 - 1.1\%. The final
systematic errors are obtained by quadratically summing the
errors due to the reconstruction efficiency and the fitting
systematics. The summary of the systematic errors is shown in Table
\ref{tab:sys}. 

The detector bias is the dominant systematic error for 
$A_{CP}$($B^+ \to \overline{K}{}^0K^+$) and 
$A_{CP}$($B^+ \to K^0\pi^+$); the systematic uncertainties evaluated from the 
partial rate asymetry of the continuum background are 0.02 and 0.01 for 
these two modes, respectively. The 
systematic errors for $A_{CP}(B^0 \to \overline{K}{}^0 K^0)$ are estimated by
varying the fitting parameters by $\pm 1 \sigma$. We include
also the errors due to
$w_k$, $\chi_d$ and tag-side interference \cite{tsit} and obtain 
a total systematic error of 0.04.

In summary, using a data sample with 449 million $B\overline{B}$ pairs, we observe $B^+\to
\overline{K}{}^0 K^+$ and $B^0\to \overline{K}{}^0 K^0$ with branching
fractions $\br(B^+\to \overline{K}{}^0 K^+) = (1.22
^{+0.32+0.13}_{-0.28-0.16})\times 10^{-6}$ and $\br(B^0\to
\overline{K}{}^0 K^0) = (0.87 ^{+0.25}_{-0.20} \pm 0.09)\times
10^{-6}$. The corresponding partial-rate asymmetries are
$A_{CP}(B^+\to \overline{K}{}^0 K^+) = 0.13^{+0.23}_{-0.24}\pm 0.02$
and $A_{CP}(B^0\to \overline{K}{}^0 K^0) = -0.58^{+0.73}_{-0.66} \pm 0.04$.
In addition, we improve the measurements of the branching
fraction and partial-rate asymmetry for the decay $B^+\to
K^0\pi^+$: $\br(B^+\to K^0\pi^+) = (22.8^{+0.8}_{-0.7}\pm
1.3)\times 10^{-6}$ and $A_{CP}(B^+\to K^0 \pi^+) = 0.03 \pm 0.03
\pm 0.01$. Our measurements are consistent with previous results
\cite{br,aub1,kk}. The new results, except for $\br(B^0\to K^+ K^-)$ and 
$A_{CP}(B^0\to \overline{K}{}^0 K^0)$, have better precision than previously measured
values. Our results agree
with some theoretical predictions \cite{plb253,plb341,pqcd_kk,ben,
fleischer}. No signal is observed for $B^0\to K^+K^-$, and
we set an upper limit of $4.1 \times 10^{-7}$ at the 90\%
confidence level using the Feldman-Cousins approach \cite{upli}.

We thank the KEKB group for excellent operation of the
accelerator, the KEK cryogenics group for efficient solenoid
operations, and the KEK computer group and
the NII for valuable computing and Super-SINET network
support.  We acknowledge support from MEXT and JSPS (Japan);
ARC and DEST (Australia); NSFC and KIP of CAS (China);
DST (India); MOEHRD, KOSEF and KRF (Korea);
KBN (Poland); MIST (Russia); ARRS (Slovenia); SNSF (Switzerland);
NSC and MOE (Taiwan); and DOE (USA).


\end{document}